\documentclass[
reprint,
secnumarabic,
amssymb, amsmath,
aps,prl,
groupedaddress,
frontmatterverbose,
]{revtex4-1}
\usepackage{amsmath}
\usepackage{graphicx}
\usepackage{docs}%
\usepackage{bm}%
\usepackage{mathtools}
\usepackage[colorlinks=true,linkcolor=blue]{hyperref}%
\expandafter\ifx\csname package@font\endcsname\relax\else
\expandafter\expandafter
\expandafter\usepackage
\expandafter\expandafter
\expandafter{\csname package@font\endcsname}%
\fi
\begin{document}

\title{Hybrid Spintronic-CMOS Spiking Neural Network With On-Chip Learning: Devices, Circuits and Systems}

\author{Abhronil Sengupta}
\email{asengup@purdue.edu}
\author{Aparajita Banerjee}
\author{Kaushik Roy}

\affiliation{School of Electrical \& Computer Engineering, Purdue University, West Lafayette, IN 47907, USA}%

\begin{abstract}
Over the past decade Spiking Neural Networks (SNN) have emerged as one of the popular architectures to emulate the brain. In SNN, information is temporally encoded and communication between neurons is accomplished by means of spikes. In such networks, spike-timing dependent plasticity mechanisms require the online programming of synapses based on the temporal information of spikes transmitted by spiking neurons. In this work, we propose a spintronic synapse with decoupled spike transmission and programming current paths. The spintronic synapse consists of a ferromagnet-heavy metal heterostructure where programming current through the heavy metal generates spin-orbit torque to modulate the device conductance. Low programming energy and fast programming times demonstrate the efficacy of the proposed device as a nanoelectronic synapse. We perform a simulation study based on an experimentally benchmarked device-simulation framework to demonstrate the interfacing of such spintronic synapses with CMOS neurons and learning circuits operating in transistor sub-threshold region to form a network of spiking neurons that can be utilized for pattern recognition problems. 
\end{abstract}

\maketitle

\section{Introduction}

Brain-inspired computing models have emerged as one of the most powerful tools for pattern recognition and classification problems over the past few decades \cite{ghosh2009spiking}. Such schemes attempt to develop abstract models of the communication and functionalities involved in the neurons and synapses in the human brain in order to construct computing tools efficient at recognition and cognitive tasks. However, implementation of such non-von Neumann computing schemes on general-purpose supercomputers have not been able to harness the energy efficiency of the human brain. The sequential fetch, decode and execute cycles involved in traditional von-Neumann computing are in complete contrast to the parallel, event driven processing involved in the mammalian cortex. For instance, the IBM $Blue$ $Brain$ project \cite{markram2006blue} utilized the Blue Gene supercomputer to simulate brain activity in animals and consumed orders of magnitude more energy than the brain, even at neuron firing rates much slower than the biological time scale.

Custom CMOS analog and digital VLSI neurocomputing platforms have been also utilized to implement neuron and synapse functionalities. The $BrainScaleS$ \cite{schemmel2008wafer}, $SpiNNaker$ \cite{jin2010modeling} and the IBM $TrueNorth$ \cite{merolla2011digital} are instances of such neurocomputers based on conventional CMOS technology. However, the significant mismatch between the neuroscience mechanisms involved in the brain and the CMOS transistors have limited the capability of such computing technologies to achieve the area or power efficiency of the brain. For example, four 8-T SRAM cells (32 CMOS transistors) are required to implement the functionality of a single 4-bit synapse in a digital CMOS implementation \cite{rajendran2013specifications}. 

Recently neurocomputing architectures based on emerging post-CMOS technologies have gained popularity as they offer a direct mapping to many of the neuroscience mechanisms involved in biological synapses \cite{jackson2013nanoscale,jo2010nanoscale,ramakrishnan2011floating,nishitani2013dynamic,kuzum2011nanoelectronic} and neurons \cite{sharad2012spin,ramasubramanian2014spindle,:/content/aip/journal/apl/106/14/10.1063/1.4917011}. In order to achieve an integration density similar to the brain, neuromorphic computing architectures aim to achieve a fan-out of 10,000 for each neuron, thereby requiring orders of magnitude more synapses than neurons. Additionally, unsupervised learning using Spike-Timing Dependent Plasticity (STDP), or other Hebbian learning rules, require online programming of synapses during spike transmission. Hence, a nanoelectronic device emulating synaptic functionalities is an essential component of spiking neuromorphic architectures.
\begin{figure}[b]
\centering
\includegraphics[width = 3.2in ]{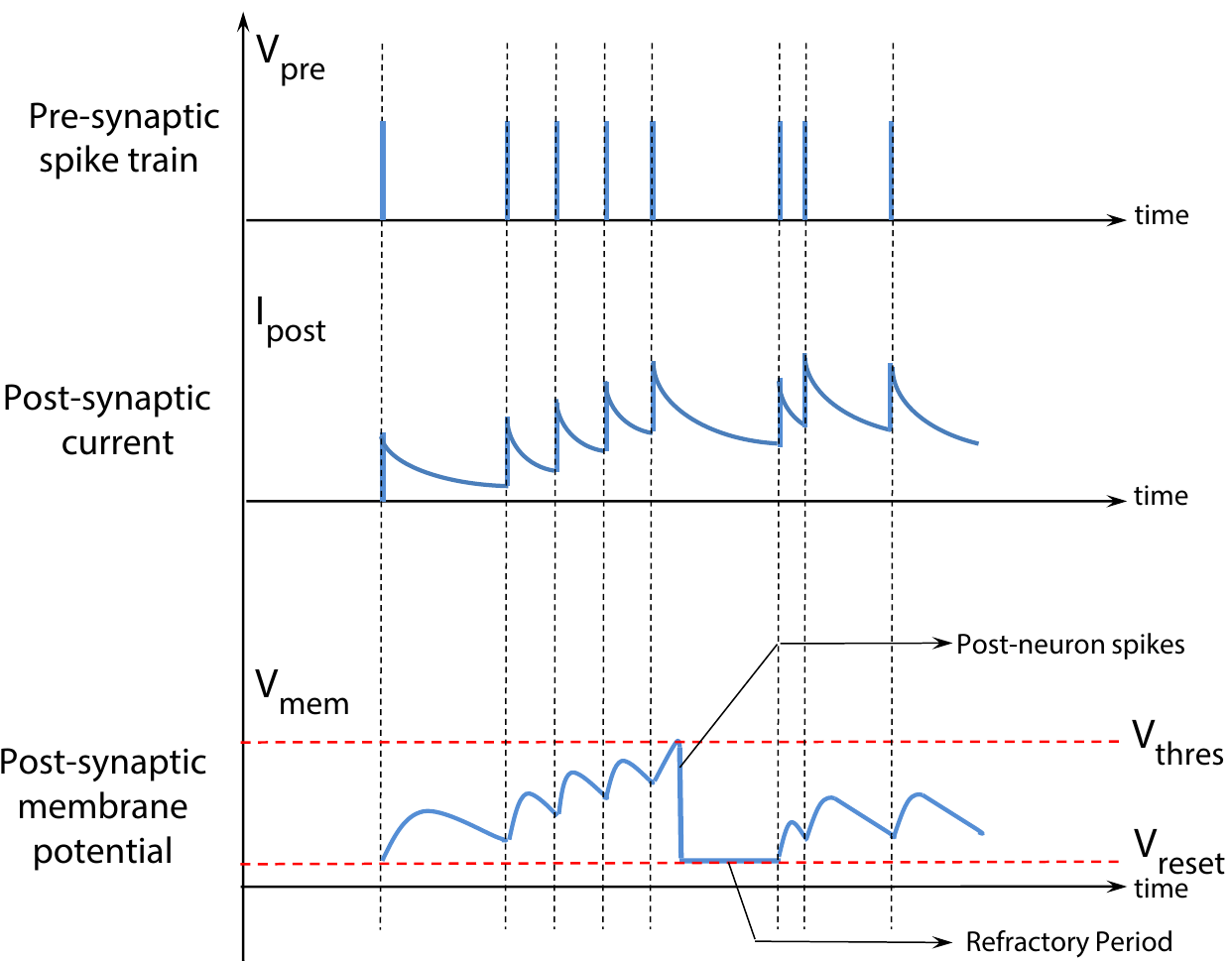}
\caption{Neuron and synapse dynamics in response to a spike train.}
\label{fig:prelims}
\end{figure}

In this work, we propose a ferromagnet (FM)-heavy metal (HM) multilayer structure where spin-orbit torque induced by the programming current flowing through the HM is the main underlying physical mechanism for generating synaptic plasticity. The ferromagnet is part of a Magnetic Tunneling Junction (MTJ) structure where spike voltage transmitted through the MTJ gets modulated by the MTJ conductance. The proposed three-terminal device structure offers the advantage of decoupled spike transmission and programming current paths thereby leading to efficient implementation of on-chip learning. Further, the proposed synapse can be programmed at low current magnitudes and small programming time durations and thereby consumes orders of magnitude lower programming energy in comparison to other state-of-the-art emerging synaptic devices. We discuss a comprehensive framework for simulating such spintronic synapse based spiking neural systems from the device (including calibration to experimental results) to the system level for performing recognition tasks. 

\section{Spiking Neural Networks: Preliminaries}\label{sec:snnbasics}

\subsection{Neuron and Synapse dynamics in Spiking Neural Networks}\label{subsec:dynamics}
A synapse is a junction connecting two neurons. The transmitting neuron is termed as the pre-neuron while the receiving neuron is termed as the post-neuron. The pre-neuron transmits a train of voltage spikes which may be represented by a set of Dirac-delta functions at time instants $t_{f}$,
\begin{equation}
V_{pre}=\sum\limits_{f} \delta(t-t_f)
\end{equation}
The synapse response to such a spike train is modelled by,
\begin{equation}
\tau_{post} \frac{dI_{post}}{dt}=-I_{post}+w\sum\limits_{f} \delta(t-t_f)
\end{equation}
where, $I_{post}$ is the post-synaptic current produced by the synapse characterized by weight $w$ and $\tau_{post}$ is the time-constant of the post-synaptic current. Hence, the post-synaptic current increases by an amount modulated by the synapse conductance (weight) at each spike instant and then starts decaying exponentially. The temporal dynamics of the leaky-integrate-fire neuron in response to such a post-synaptic current is given by,
\begin{equation}
\tau \frac{dV_{mem}}{dt}=-V_{mem}+R_{mem}\sum\limits_{i} I_{post,i}
\end{equation}
where, $V_{mem}$ is the membrane potential, $R_{mem}$ is the membrane resistance, $I_{post,i}$ is the post-synaptic current input from the $i$-th neuron, and $\tau$ is the membrane time-constant. Fig. \ref{fig:prelims} shows the temporal characteristics of the neuron and synapse in response to a series of voltage spikes transmitted from the pre-neuron. When the neuron's membrane potential $V_{mem}$ crosses the threshold $V_{thres}$, the membrane potential gets reset to $V_{reset}$ and does not vary for a time duration termed as the refractory period.

\subsection{Learning: STDP}\label{subsec:STDP}
\begin{figure}
\centering
\includegraphics[width = 2.9in ]{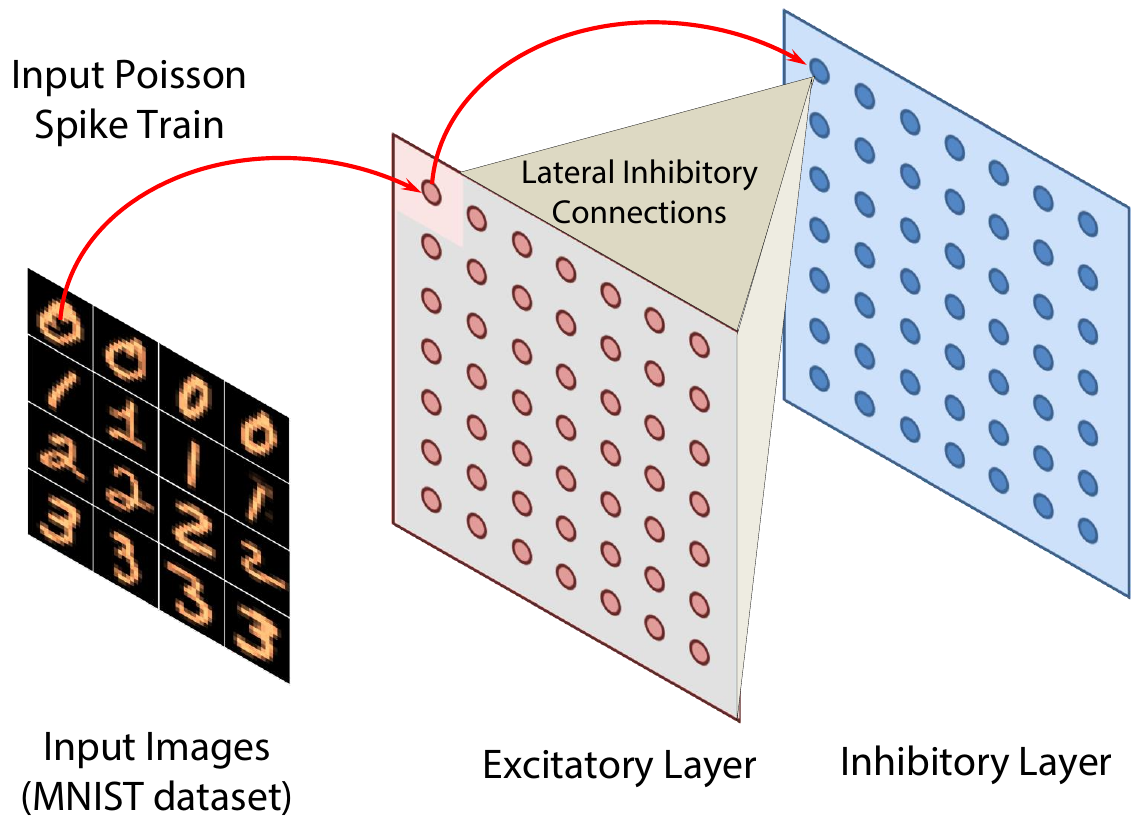}
\caption{Network connectivity utilized for pattern recognition. Neurons with lateral inhibitory connections receive input Poisson spike trains with average rate proportional to pixel intensity. }
\label{fig:network}
\end{figure}
According to the theory of Hebbian Learning \cite{morris1999hebb}, synaptic weight or conductance is modulated depending on the spiking patterns of the pre-neuron and post-neuron. STDP, a form of Hebbian learning, states that the weight of the synapse increases (decreases) if the pre-neuron spikes before (after) the post-neuron. Intuitively, this signifies that the synapse strength should increase if the pre-neuron spikes before the post-neuron as the pre-neuron and post-neuron appear to be temporally correlated. The relative change in synaptic strength decreases exponentially with the timing difference between the pre-neuron and post-neuron spikes. The STDP characteristics have been formulated in a mathematical framework based on measurements for rat hippocampal glutamatergic synapses \cite{bi2001synaptic},
\begin{equation}
\begin{aligned}
\Delta w &=A_{+}\exp\left(\frac{-\Delta t}{\tau_{+}}\right), \Delta t > 0\\
&=-A_{-}\exp\left(\frac{\Delta t}{\tau_{-}}\right), \Delta t < 0
\end{aligned}
\end{equation}
Here, $A_{+}, A_{-}, \tau_{+}$ and $\tau_{-}$ are constants and $\Delta t = t_{post}-t_{pre}$, where $t_{pre}$ and $t_{post}$ are the time-instants of pre- and post-synaptic firings respectively. We will refer to the case of $\Delta t >0$ ($\Delta t <0$) as the positive (negative) time window for learning.
\subsection{Spike Frequency Adaptation}

In order to model spike frequency adaptation mechanisms observed in biological neurons, an additional slowly varying adaptation parameter $a$ is introduced in the temporal dynamics of the neuron as,
\begin{equation}
\tau \frac{V_{mem}}{dt}=-V_{mem} (1+a)+R_{mem}\sum\limits_{i} I_{post,i}
\end{equation}
The adaptation parameter $a$ increases every time the neuron spikes, otherwise it decays exponentially. This implies that in case a neuron starts spiking at a high frequency, the leak parameter starts to increase to reduce its spike frequency.
\subsection{Network Connectivity}

Fig. \ref{fig:network} shows the network connectivity of spiking neurons utilized for pattern recognition problems. Such a network topology has been shown to be efficient in several pattern recognition problems like digit recognition \cite{peter2015unsup} and sparse encoding \cite{knagsparse}. Input image pixels are encoded as Poisson spike trains with average rate directly proportional to the pixel intensity. These input spike trains are received by all neurons in an excitatory layer through synapses whose weights are learnt using STDP. Each neuron in the excitatory layer is connected to a corresponding neuron in an inhibitory layer such that a spike in the excitatory neuron triggers a spike in the corresponding neuron in the inhibitory layer. Each neuron in the inhibitory layer is connected to all neurons in the excitatory layer except the neuron from which it received the input. This connectivity helps to implement lateral inhibitory connections in the excitatory layer such that when one neuron starts to spike in response to some input pattern, it prohibits the other neurons from spiking. However, in order to prevent a particular neuron from dominating the spiking pattern due to lateral inhibitory connections, spike frequency adaptation mechanism is also implemented in each neuron. The neurons in the excitatory layer are assigned classes based on their highest response (spike frequency) to input training patterns.

\section{Spintronic Synapse}\label{sec:devices}

\subsection{Spin-orbit torque driven motion of Dzyaloshinskii domain walls}\label{subsec:physics}

In this section we provide a brief discussion on the underlying physical phenomena involved in current induced domain wall motion in heavy metal (HM) - ferromagnet (FM) - insulator (I) multilayer structures.

Recent experiments on magnetic nanostrips of Pt/CoFe/MgO and Ta/CoFe/MgO have revealed high domain wall velocities due to charge current densities that are two orders of magnitude lower than that achievable by conventional spin-transfer torque (STT) \cite{emori2013current}. Additionally, domain wall motion was observed to be against the direction of electron flow (i.e. in the direction of current flow) in multilayer structures with Pt as the underlayer, thereby suggesting that current induced spin-orbit torque is the main mechanism of domain wall motion in such multilayer structures (with negligible contribution from conventional STT) \cite{emori2013current}. In such magnetic heterostructures with high perpendicular magnetocrsytalline anisotropy (PMA), spin orbit coupling and broken inversion symmetry leads to the stabilization of homochiral domain walls through the Dzyaloshinskii-Moriya exchange interaction (DMI) \cite{chen2013novel}. We restrict our analysis for Pt/CoFe/MgO multilayer structures for this text due to the possibilities of achieving high domain wall velocities ($\sim 400m/s$) \cite{martinez2014current,emori2014spin,perez2014micromagnetic}. However, the analysis can be easily extended to other magnetic heterostructures with different underlayers.

Such interfacial DMI at the FM-HM interface leads to the formation of a N\'{e}el domain wall with left-handed chirality for Pt/CoFe/MgO multilayer structures \cite{emori2013current, martinez2014current,emori2014spin,perez2014micromagnetic}. The DMI strength in such structures with HM underlayers has been observed to be sufficiently strong to impose a N\'{e}el wall configuration in FMs where conventional magnetostatics would have yielded a Bloch configuration \cite{emori2013current}. When an in-plane charge current is injected through the HM, a transverse spin-current is generated due to deflection of opposite spin-polarizations on the top and bottom surfaces of the HM. This phenomena is termed as spin-Hall effect \cite{hirsch1999spin} and arises as a consequence of spin-orbit torque. The accumulated spins at the FM-HM interface leads to DMI stabilized N\'{e}el domain wall motion. The direction of domain wall motion is in the direction of charge current flow and the final magnetization of the ferromagnet is given by the cross-product of the direction of injected spins at the FM-HM interface and the magnetization direction of the FM at the domain wall location.
\subsection{Device proposal for spintronic synapse}\label{subsec:bit-cell}

\begin{figure}
\centering
\includegraphics[width = 3.2in ]{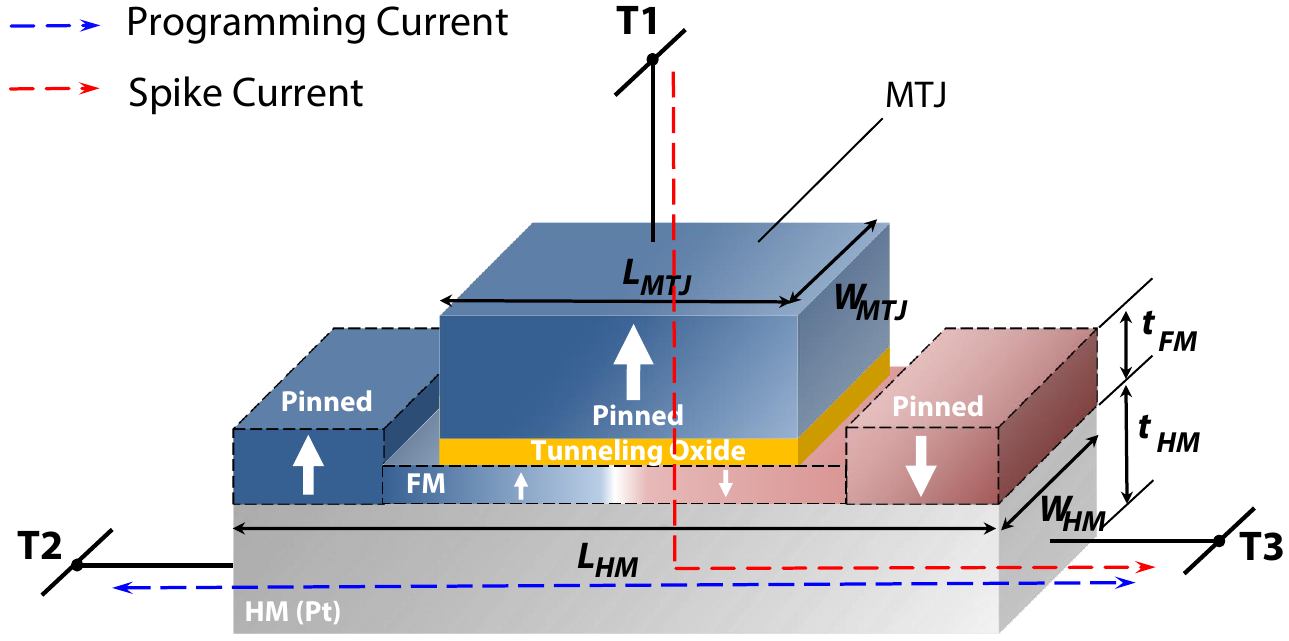}
\caption{Device structure for a spintronic synapse with decoupled spike transmission and programming current paths. Spike current flows through the MTJ structure between terminals T1 and T3. Programming current flows through the HM between terminals T2 and T3.}
\label{fig:device}
\end{figure}

Such spin-orbit torque driven domain wall motion in FMs due to charge current flow through a HM underlayer leads to the possibility of a device structure that can manifest decoupled spike transmission (read) and programming (write) current paths. We propose a three-terminal device structure consisting of a FM lying on top of a HM (Fig. \ref{fig:device}). The FM is part of an MTJ structure where the FM is separated from a Pinned layer (magnetic region whose magnetization is fixed) by a Tunneling Oxide barrier (MgO). The FM has two additional Pinned layers on either side to ensure that the domain wall stabilizes at the extreme locations of the FM for sufficiently large values of the programming current. While the spike current flows through the MTJ structure between terminals T1 and T3, the programming current flows through the HM layer between terminals T2 and T3. Note that a preliminary synaptic device proposal based on Bloch domain wall motion due to spin-orbit torque was explored previously in Ref. \cite{sengupta2016spin}. However, an external magnetic field was required to modulate the device conductance during learning. Further the magnet width was not scalable beyond $100nm$ to ensure Bloch wall orientation. The current device proposal based on N\'{e}el wall motion is not only more energy efficient, but also requires no external magnetic field for domain wall motion due to the inherent interfacial DMI. Further this work provides a synergistic device-circuit-system perspective for the implementation of STDP in SNNs utilizing the proposed spintronic device as the core building block.

The location of the domain wall in the FM encodes the resistance of the device lying in the path of the spike current between terminals T1 and T3 and thereby implements the synaptic functionality. On the other hand, the programming current path is completely decoupled (between terminals T2 and T3) and the resistance in the path of the programming current is mainly determined by the HM resistance. It is worth noting here that although some amount of spike current will flow through the HM, the magnitude of this current can be maintained to sufficiently low values below the domain wall depinning current since the synapses are required to drive CMOS neurons operating in the subthreshold regime. 

\subsection{Synaptic plasticity mechanism}\label{subsec:stdp}

Programming current flowing from terminal T2 to terminal T3 results in domain wall motion in the same direction so that the +z domain in the FM starts to expand and vice versa. For a given duration of the programming current pulse, the domain wall displacement is directly proportional to the magnitude of the programming current.

On the other hand, the device conductance between terminals T1 and T3 varies linearly with the domain wall position. Let us denote the conductance of the device when the entire FM magnetization is parallel (anti-parallel) to the Pinned layer as $G_{P} (G_{AP})$, i.e. the domain wall is at the extreme right (left) of the FM. Thus, for an intermediate position of the domain wall at a position $x$ from the left-edge of the MTJ, the device conductance between terminals T1 and T3 is given by,

\begin{equation}
G_{eq} = G_{P}.\frac{x}{L} + G_{AP}. \left(1 - \frac{x}{L}\right) + G_{DW} 
\label{eq:res_model}
\end{equation}
where, $L$ denotes the length of the MTJ excluding the domain wall width and $G_{DW}$ represents the conductance of the wall region. It is worth noting here, that $L, G_{DW}, G_{P}$ and $G_{AP}$ are all constants (for constant voltage drop across the MTJ). Due to such a linear relationship between domain wall position and device conductance, the programming current is directly proportional to the change in device conductance (which encodes the synaptic weight) for a fixed duration of the programming signal.

\subsection{Spiking neuromorphic architecture based on spintronic synapse}\label{subsec:architecture}

\begin{figure}[b!]
\centering
\includegraphics[width = 3.2in ]{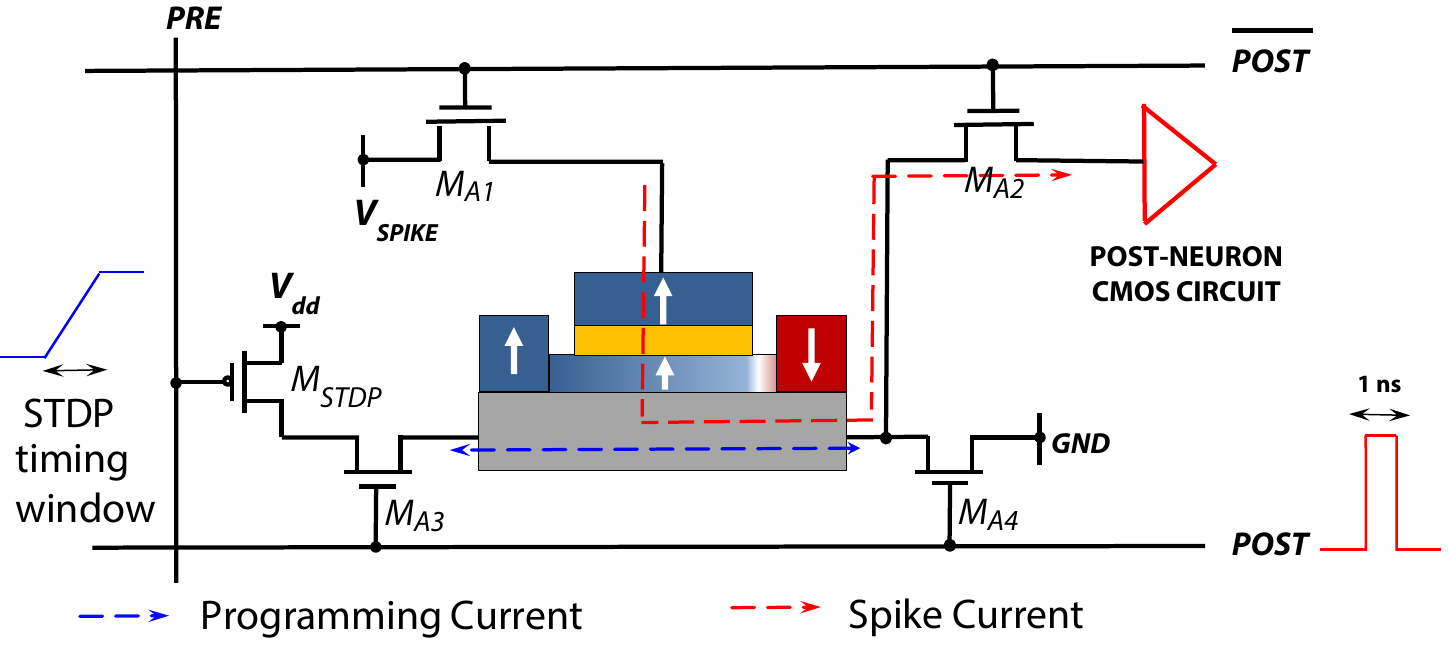}
\caption{Spintronic synapse with access transistors to decouple the programming and spike current paths.}
\label{fig:bit_cell}
\end{figure}

Fig. \ref{fig:bit_cell} represents possible arrangement of a spintronic synapse with access transistors $M_{A1}-M_{A4}$ to decouple the programming and spike current paths. The access transistors act as switches to select the appropriate terminals of operation for the device. The operating mode of the synapse, i.e. the spike transmission mode or programming mode is accomplished by the control signal POST. The POST signal is activated during the programming mode of operation of the synapse. 

The PRE line is used to pass the necessary amount of programming current required for the corresponding weight change involved due to the delay between the pre-neuron and post-neuron spikes. A negative (positive) current should flow through the HM for the negative (positive) time window duration. Since the programming current amplitude is directly proportional to the amount of weight change, the current signal flowing through the HM should vary in a similar fashion as the STDP learning curve (exponentially) with the time delay between the pre-neuron and post-neuron spikes. 

For simplicity, let us discuss the case for the positive time window. The exponential variation of current through the HM can be obtained by a transistor operating in sub-threshold regime since the current flowing through the transistor will vary exponentially with the gate to source voltage. Thus for a linear increase of voltage of the PRE line with time, the transistor $M_{STDP}$ will be driven from cut-off to saturation regime when the POST signal is activated and an appropriate programming current should flow through the HM. It is worth noting here that the HM resistance $\sim$ a few hundred ohms and the maximum programming current required is $\sim$ a few tens of $\mu A$, thereby leading to a very small voltage drop across the device when the POST signal is activated. Fig. \ref{fig:bit_cell} shows the interface circuits involved in the synapse programming for the positive time window. A similar approach can be adopted to program the synapses for the negative time window (by utilizing an NMOS operating in sub-threshold saturation driven by a linearly increasing gate voltage to pass programming current from terminals T3 to T2) and the two learning circuits for the negative and positive timing windows have to be activated sequentially everytime the pre-neuron spikes. Since the time duration involved in programming is $\sim$ a few $ns$ in comparison to learning time constants used in this work $\sim \mu s$, the POST signal essentially samples the necessary amount of programming current from the PRE line (programming current magnitude determined by $M_{STDP}$ transistor). 

In our proposed programming scheme, we program the synapses only when the post-neuron spikes. Hence, in order to account for the negative and positive time windows involved in STDP learning, the POST signal should be activated with a delay corresponding to the time duration of the negative timing window in order to sample the programming current contributions from the learning circuits for both the timing windows. 

\begin{figure}[t]
\centering
\includegraphics[width = 3.4in ]{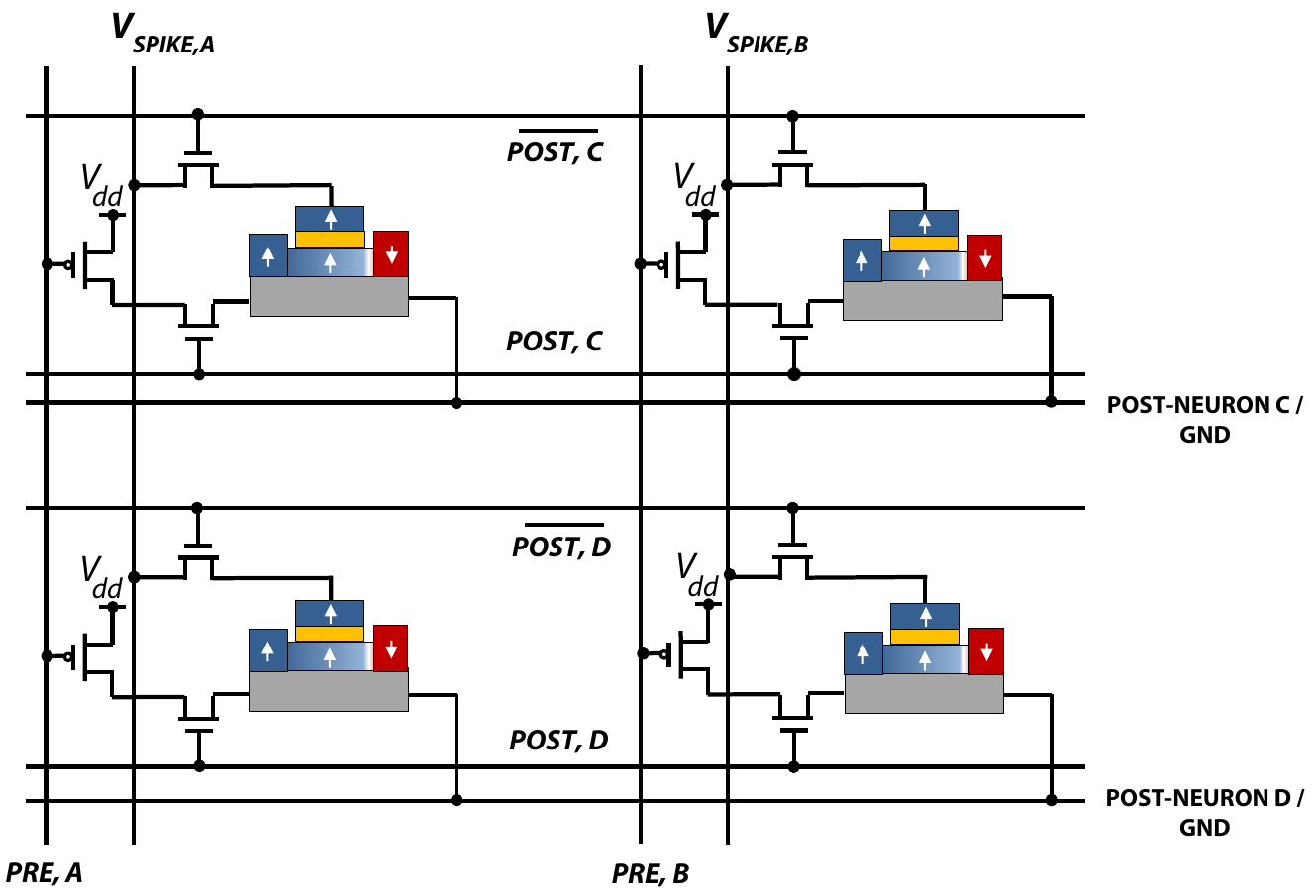}
\caption{Possible arrangement of synapses in an array interfaced with CMOS neurons and programming circuits. The figure shows synapses connecting pre-neurons A and B to post-neurons C and D.}
\label{fig:array}
\end{figure}

Arrangement of synapses in an array fashion as shown in Fig. \ref{fig:array}, interfaced with CMOS neurons can lead to dense spiking neuromorphic architectures. Please note that the access transistors $M_{A2}$ and $M_{A4}$ for terminal T3 of the device (Fig. \ref{fig:bit_cell}) can be shared across the row such that the corresponding horizontal line connecting terminals T3 for the devices in a particular row are driven to GND (POST signal is HIGH) or the post-neuron circuit (POST signal is LOW). Details of the CMOS circuits involved in the programming scheme and neuron implementation will be discussed in the next section.

\section{CMOS Learning and Neuron Circuits}\label{sec:circuits}

\subsection{Sub-threshold circuit for STDP learning}\label{subsec:stdp}
\begin{figure}[t]
\centering
\includegraphics[width = 2.4in ]{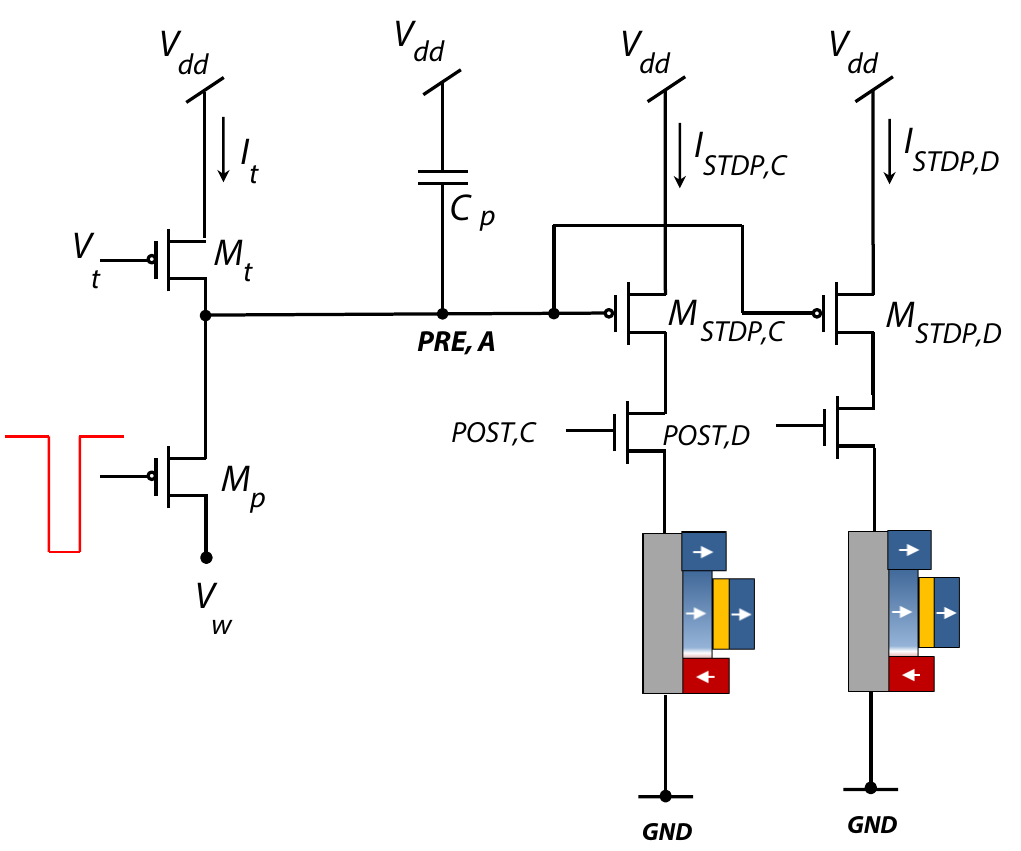}
\caption{Sub-threshold CMOS circuit utilized for generating the programming current involved in STDP learning (circuit for positive time window shown) for pre-neuron A connecting to post-neurons C and D.}
\label{fig:prog_circuit}
\end{figure}
\begin{figure*}[t]
\centering
\includegraphics[width = 7.0in ]{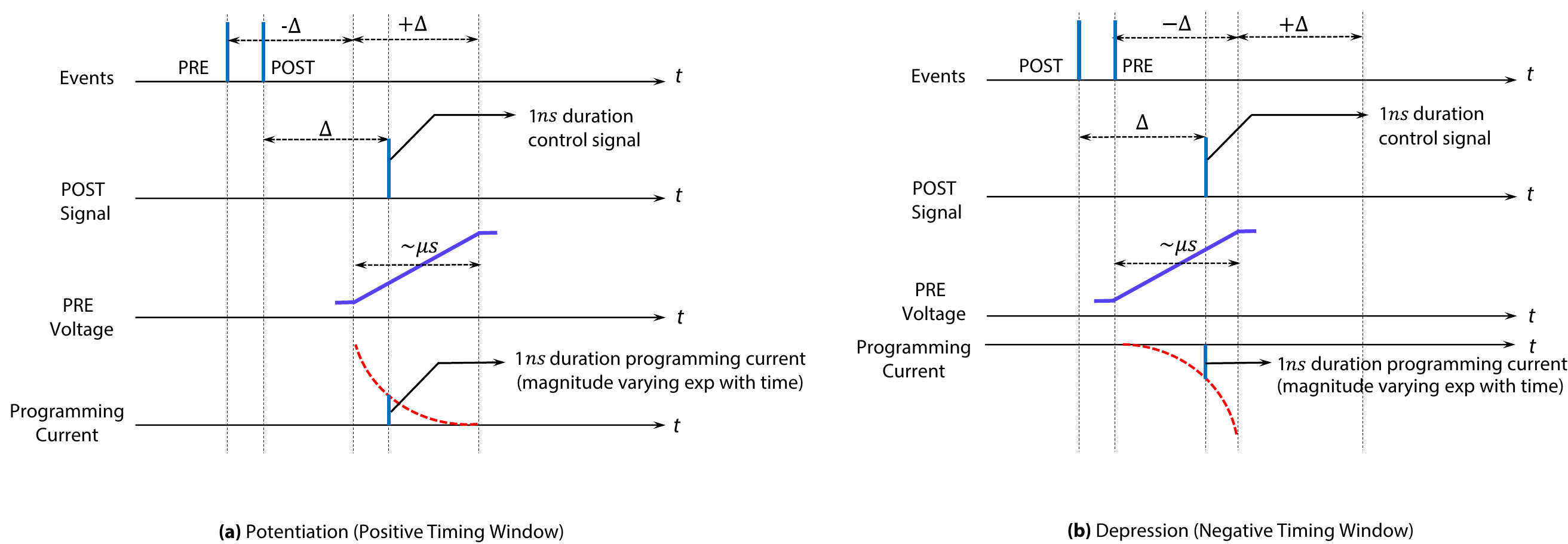}
\caption{Detailed timing diagrams demonstrating the implementation of (a) potentiation (positive timing window) and (b) depression (negative timing window) in the spintronic synapse. POST is the control signal that is activated during programming while PRE is the gate voltage of the $M_{STDP}$ transistor that implements synaptic plasticity. Duration of the programming current is determined by the duration of the POST signal while the magnitude is determined by the value of the PRE signal when the POST signal is high.}
\label{pulse_seq}
\end{figure*}
The circuit involved in generating the PRE signal is discussed in this section. Fig. \ref{fig:prog_circuit} shows the sub-threshold CMOS circuit used to generate the PRE signal for pre-neuron A connecting to post-neurons C and D. We discuss the mechanism for generating the signal for the positive time window. A similar design can be used to generate the programming current for the negative time window. The circuit was originally proposed in \cite{lazzaro1994low} as a reset and discharge synapse. However it failed to emulate the post-synaptic dynamics of biological synapses as the circuit response depends only on the previous input spike \cite{bartolozzi2007synaptic}. In this work, we employ this circuit to implement STDP learning in our proposed device.

The transistor $M_{p}$ acts as a switch. When the positive time window starts, the transistor $M_{p}$ receives a low-active pulse and gets turned ON. As a result, the node PRE, A is set to the bias voltage $V_{w}$. After the transistor $M_p$ is switched OFF, the transistor $M_{t}$, operating in sub-threshold saturation regime, provides a constant current to linearly charge the capacitor $C_p$ at a rate $\frac{I_t}{C_p}$. Hence, if the transistor $M_{STDP}$ is operated in sub-threshold saturation, exponential dynamics will be observed in the output current $I_{STDP}$. The current flowing through transistor $M_{STDP}$ for an input pulse at time $t=t_n$ is given by,
\begin{equation}
I_{STDP} = I_{0}e^{\frac{-U_{T} C_p (t-t_n)}{\emph{k}I_t}}
\label{eq:stdp}
\end{equation}
where, $\emph{k}$ is the sub-threshold slope factor and $U_{T}$ is the thermal voltage. Hence, whenever the pre-neuron spikes, the circuits for generating the STDP characteristics for the negative and positive time windows are activated sequentially. When learning starts for the positive timing window, a short pulse is applied to the gate of the transistor $M_p$ so that the circuit is reset and the node PRE, A is charged to $V_w$. When the post-neuron does not spike, the transistor $M_{STDP}$ is in cut-off since the POST signal is deactivated and the access transistors for programming are turned OFF. Once the post-neuron spikes, the programming current path gets activated and the transistor $M_{STDP}$ switches to the sub-threshold saturation regime and transmits the necessary amount of programming current through the device. Note that apart from the transistor $M_{STDP}$ (one transistor for each of the positive and negative timing windows), the entire learning circuitry can be shared across the column of the crossbar array.

The operation is discussed in details in Fig. \ref{pulse_seq}. Let us first describe the case for the positive timing window, i.e. post-neuron spiking after the pre-neuron (Fig. \ref{pulse_seq}(a)). ($-\Delta$)/($+\Delta$) represents the duration during which the learning circuits for the negative/positive timing windows are activated sequentially for the corresponding pre-neuronal firing event. The control signal POST is activated after a duration ($\Delta$) the post-neuron spikes. As described in the figure, magnitude of the programming pulse is determined by the current being passed by the programming transistor $M_{STDP}$ (value of the PRE voltage when the POST signal is active) and the duration is determined by the duration of the POST signal. Since the PRE signal varies in $\sim \mu s$ time scale and does not almost change during the programming time duration ($\sim ns$ time scale), it ensures that the programming current magnitude is almost constant and is equal to the sampled value from the exponential STDP dynamics corresponding to the appropriate spike timing difference. As mentioned previously, since the programming current magnitude is directly proportional to the amount of change in the MTJ conductance, exponential STDP characteristics is implemented in the spintronic device. Similar discussions are valid for the negative timing window (Fig. \ref{pulse_seq}(b)) where the post-neuron spikes before the pre-neuron. In this case, the POST signal is activated during the negative window ($-\Delta$) and the NMOS transistor passes an appropriate amount of programming current in the opposite direction through the device. Circuit-level simulations confirming the proposal have been demonstrated in Fig. \ref{fig:negf_stdp}(b).
\subsection{Differential Pair Integrator circuit for post-synaptic current generation}\label{subsec:dpi}

\begin{figure*}[t]
\centering
\includegraphics[width = 6.6in ]{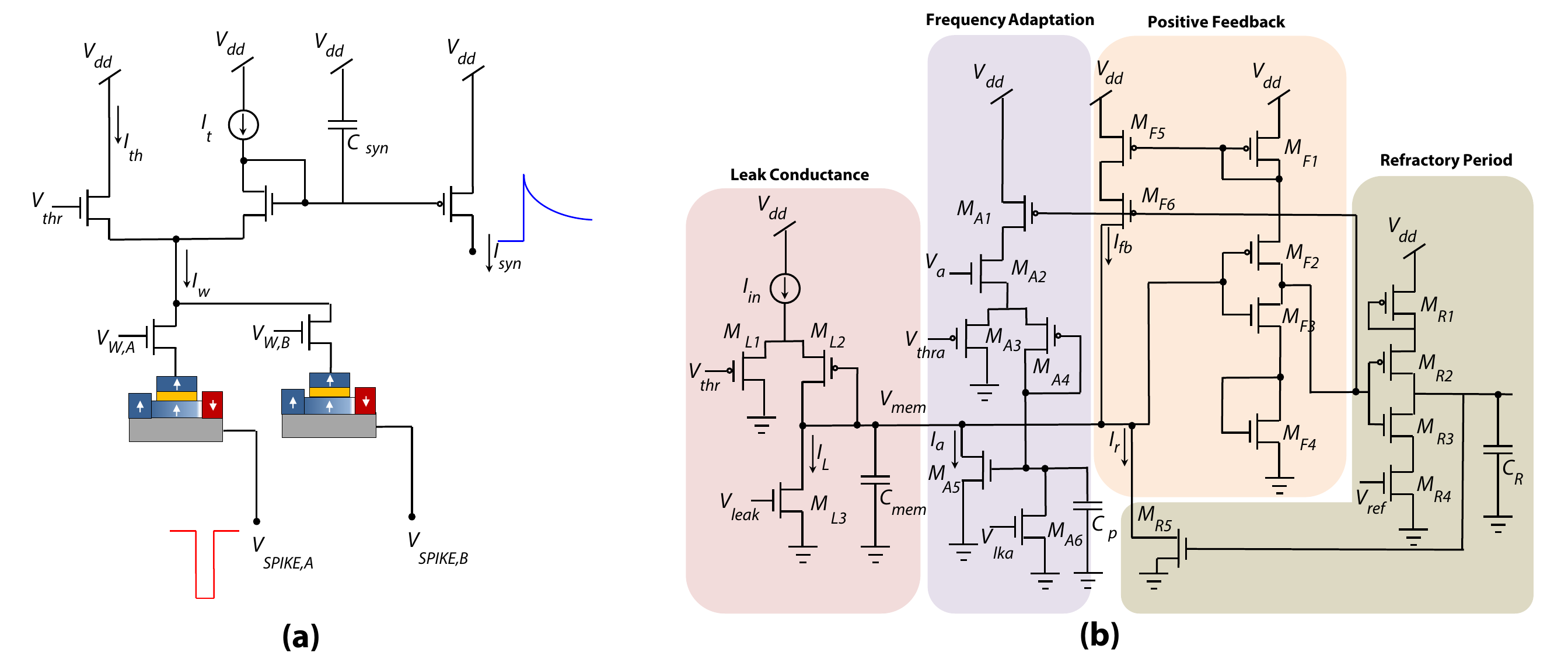}
\caption{(a) DPI circuit interfaced with spintronic synapses to emulate synaptic dynamics. (b) Sub-threshold CMOS neuron with leak conductance, spike frequency adaptation, positive feedback and refractory period implementation blocks \cite{chicca2014neuromorphic}.}
\label{fig:postsyn_circuit}
\end{figure*}
The Differential Pair Integrator (DPI) circuit has been a popular mechanism for generating synaptic dynamics \cite{chicca2014neuromorphic} and integration of such DPI circuits with memristor synapses has been recently proposed \cite{indiveri2013integration}. Fig. \ref{fig:postsyn_circuit}(a) shows how such DPI circuits can be integrated with our proposed spintronic synapses to generate exponential post-synaptic currents in response to input spikes. Assuming all transistors are in sub-threshold saturation and using the translinear principle \cite{chicca2014neuromorphic, indiveri2013integration} it can be shown that the output current $I_{syn}$ exhibits temporal dynamics of the form,
\begin{equation}
\tau_{syn} \frac{d I_{syn}}{dt}+I_{syn} = \frac{I_w I_{th}}{I_t}
\label{eq:dpi}
\end{equation}
where, $\tau=\frac{CU_T}{\emph{k}I_t}$. The above relationship is valid if the circuit is operated in the linear region ($I_t\ll I_w$). The bias voltage $V_w$ acts as a scaling gain factor for the post-synaptic current. On the arrival of an input spike, the current $I_w$ gets modulated by the MTJ conductance and thereby causes $I_{syn}$ to increase by an amount governed by the synaptic weight. When there is no spike transmission, $I_{syn}$ decreases exponentially thereby emulating the synaptic dynamics discussed earlier. The access transistors driven by $\overline{\textrm{POST}}$ signal have not been shown in Fig. \ref{fig:postsyn_circuit} but are present in the design to ensure that the programming current path is deactivated when spike transmission path is enabled.

\subsection{Sub-threshold CMOS neuron}\label{subsec:neuron}

CMOS circuits operating in sub-threshold (Fig. \ref{fig:postsyn_circuit}(b)) have been shown to replicate a wide range of temporal dynamics observed in biological neurons like spike frequency adaptation and refractory period generation \cite{chicca2014neuromorphic,indiveri2003low,livi2009current}. When operated in the sub-threshold regime, the main mechanism of carrier transport in CMOS transistors is diffusion, thereby emulating the mechanism of ion flow in biological neuron channels \cite{chicca2014neuromorphic}. 

$I_{in}$ represents the input current provided to the neuron. Using the translinear principle and assuming all transistors in sub-threshold saturation, it can be shown that the temporal dynamics of $I_{mem}$ is given by \cite{chicca2014neuromorphic},
\begin{equation}
\tau_{mem} \frac{d I_{mem}}{dt}+I_{mem}\left(1+\frac{I_{a}}{I_t}\right) = \frac{I_{in} I_{th}}{I_t}
\label{eq:dpi}
\end{equation}
where, $\tau=\frac{C_{mem}U_T}{\emph{k}I_t}$. The above relation is again valid when the DPI circuit operates in the linear region (i.e. $I_t\ll I_{in}$).

We would like to conclude this section by relating the computing models discussed in Section II to circuit implementations discussed in Section IV. Postsynaptic and neuron dynamics (referred in Eqs. 2 and 5) can be directly mapped to the DPI circuit and subthreshold CMOS neuron circuit (referred in Eqs. 8 and 9) respectively. Readers are referred to Ref. \cite{ chicca2014neuromorphic} for details on neuromorphic chips utilizing such analog CMOS neurons and interfacing such circuits with post-CMOS synaptic crossbar arrays. Our proposal in this work includes the implementation of plasticity mechanism (referred in Eq. 4) in the spintronic device structure utilizing the device concepts (presented in Section III) and learning circuit primitives (presented in Section IV.1).

\section{Simulation Results}\label{sec:simulation}

\subsection{Simulation Framework}\label{subsec:framework}

In order to simulate the SNN implementation based on the proposed spintronic synapse, a hierarchical simulation framework was utilized. Device-level simulations of the spin-orbit torque induced domain wall motion was performed in MuMax \cite{vansteenkiste2011mumax}, a GPU accelerated micromagnetic simulation tool. A behavioral model of the device was developed for subsequent simulation of such synapses interfaced with CMOS neurons and learning circuits. The circuit level simulations were performed in HSPICE using a standard cell library in commercial 45nm CMOS technology. The device and circuit simulations were utilized to generate models of the plastic synapses and spiking neurons to perform system level simulations of a network of spiking neurons using Brian simulator \cite{goodman2009brian}. 

\subsection{Device Level Simulations}\label{subsec:device_simulation}

\begin{figure*}
\centering
\includegraphics[width = 5in ]{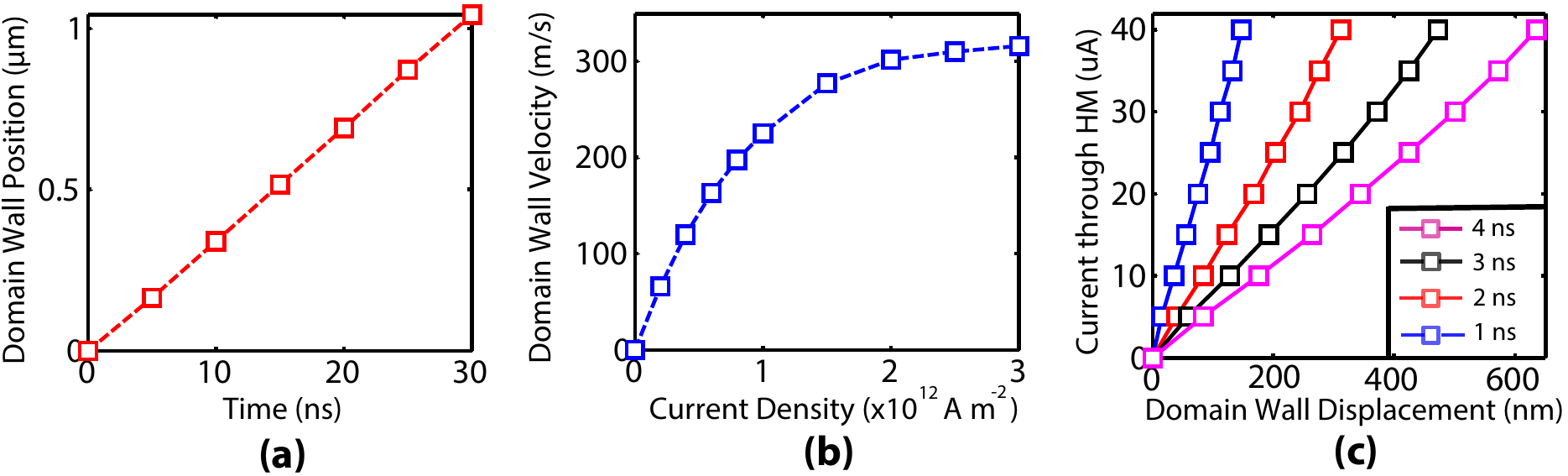}
\caption{(a) Domain wall displacement as a function of time for a CoFe strip of cross-section $160nm \times 0.6nm$ due to the application of a charge current density, $J= 0.1 \times 10^{12} A m^{-2}$. (b) Domain wall velocity as a function of current density. The results are in good agreement with \cite{martinez2014current}. (c) Domain wall displacement is directly proportional to the programming current for a fixed duration of the programming pulse.}
\label{fig:calib}
\end{figure*}

The magnetization dynamics of the ferromagnet can be described by solving Landau-Lifshitz-Gilbert equation with additional term to account for the spin-orbit torque generated by spin-Hall effect at the FM-HM interface ~\cite{martinez2014current,slonczewski1989conductance},
\begin{equation}
\frac {d\widehat {\textbf {m}}} {dt} = -\gamma(\widehat {\textbf {m}} \times \textbf {H}_{eff})+ \alpha (\widehat {\textbf {m}} \times \frac {d\widehat {\textbf {m}}} {dt})+\beta (\widehat {\textbf {m}} \times \widehat {\textbf {m}}_P \times \widehat {\textbf {m}})
\end{equation}
where, $\widehat {\textbf {m}}$ is the unit vector of FM magnetization at each grid point, $\gamma= \frac {2 \mu _B \mu_0} {\hbar}$ is the gyromagnetic ratio for electron, $\alpha$ is Gilbert\textquoteright s damping ratio, $\textbf{H}_{eff}$ is the effective magnetic field, $\beta=\frac{\hbar \theta J}{2 \mu_0 e t M_s}$ ($\hbar$ is Planck’s constant, $J$ is input charge current density, $\theta$ is spin-Hall angle~\cite{martinez2014current}, $\mu_0$ is permeability of vacuum, $e$ is electronic charge, $t$ is FL thickness and $M_s$ is saturation magnetization) and $\widehat {\textbf {m}}_P$ is direction of input spin current. The effective field $\textbf{H}_{eff}$ also includes the field due to DMI and is given by,
\begin{equation}
\textbf{H}_{DMI} = -\frac{2D}{\mu_{0}M_{s}}\left[\frac{\partial m_{z}}{\partial x}\widehat{x} + \frac{\partial m_{z}}{\partial y}\widehat{y} - \left(\frac{\partial m_{x}}{\partial x}+\frac{\partial m_{y}}{\partial y}\right )\widehat{z}\right ]
\end{equation}

Here, $D$ represents the effective DMI constant and determines the strength of DMI field in such multilayer structures. A positive sign of $D$ implies right-handed chirality and vice versa. In the presence of DMI, the boundary conditions at the edges of the sample is given by,
\begin{equation}
\frac{\partial \widehat {\textbf {m}}}{\partial n} = \frac{D}{2A}\widehat {\textbf {m}} \times \left(\widehat {\textbf {n}} \times \widehat{z} \right )
\end{equation}
where, $A$ is the exchange correlation constant and $\widehat {\textbf {n}}$ represents the unit vector normal to the surface of the FM. The simulation parameters are given in Table I and was used for the rest of this work, unless otherwise stated. The parameters were obtained experimentally from magnetometric measurements of Ta(3nm)/Pt(3nm)/CoFe(0.6nm)/MgO(1.8nm)/Ta(2nm) nanostrips \cite{emori2014spin}. Current density was estimated by assuming that the current flow is mainly through the FM-HM layers in the stack structure \cite{emori2014spin}.

\begin{table}[h]
\renewcommand{\arraystretch}{1.3}
\caption{Device Simulation Parameters}
\label{table_1}
\centering
\begin{tabular}{c c}
\hline 
\bfseries { Parameters} & \bfseries { Value}\\
\hline
{ Ferromagnet Dimensions} & { $320 \times 20 \times 0.6 nm^3$} \\
{ Grid Size} & { $ 4 \times 1 \times 0.6 nm^3$} \\
{ Heavy Metal Thickness} & { $ 3 nm$} \\
{ Domain Wall Width} & { $ 7.6 nm$} \\
{ Saturation Magnetization, $M_s$} & { 700 $KA/m$} \\
{ Spin-Hall Angle, $\theta$} & { 0.07} \\
{ Gilbert Damping Factor, $\alpha$} & { 0.3} \\
{ Exchange Correlation Constant, $A$} & { $1 \times 10^{-11} J/m$} \\
{ Perpendicular Magnetic Anisotropy} & { $4.8 \times 10^{5} J/m^{3}$} \\
{ Effective DMI constant, $D$} & { $-1.2 \times 10^{-3} J/m^{2}$} \\
\hline
\end{tabular}
\end{table}

Fig. \ref{fig:calib}(a) shows the domain wall displacement in a CoFe sample with cross-section of $160nm \times 0.6nm$ for a charge current density of $J= 0.1 \times 10^{12} A/m^2$. The grid size was taken to be $ 4 \times 4 \times 0.6 nm^3$. Fig. \ref{fig:calib}(b) depicts the variation of the domain wall velocity with input charge current density. The velocity increases linearly with the current density and ultimately reaches a saturation velocity. The graphs are in good agreement with results illustrated in \cite{martinez2014current} for the same multilayer structure described in this section. Fig. \ref{fig:calib}(c) illustrates the fact that the domain wall displacement is directly proportional to the magnitude of the programming current (for domain wall velocities below the saturation regime). For a duration of $1ns$, a maximum current of $\sim 80 \mu A$ is required to displace the domain wall from one edge of the FM to the other edge. %Hence, the maximum amount of energy dissipated to program the synapses from the OFF state to the ON state (or vice versa) is $\sim 0.24fJ$ ($I^2 \times R \times t$ energy dissipation).

\begin{figure}
\centering
\includegraphics[width = 3in ]{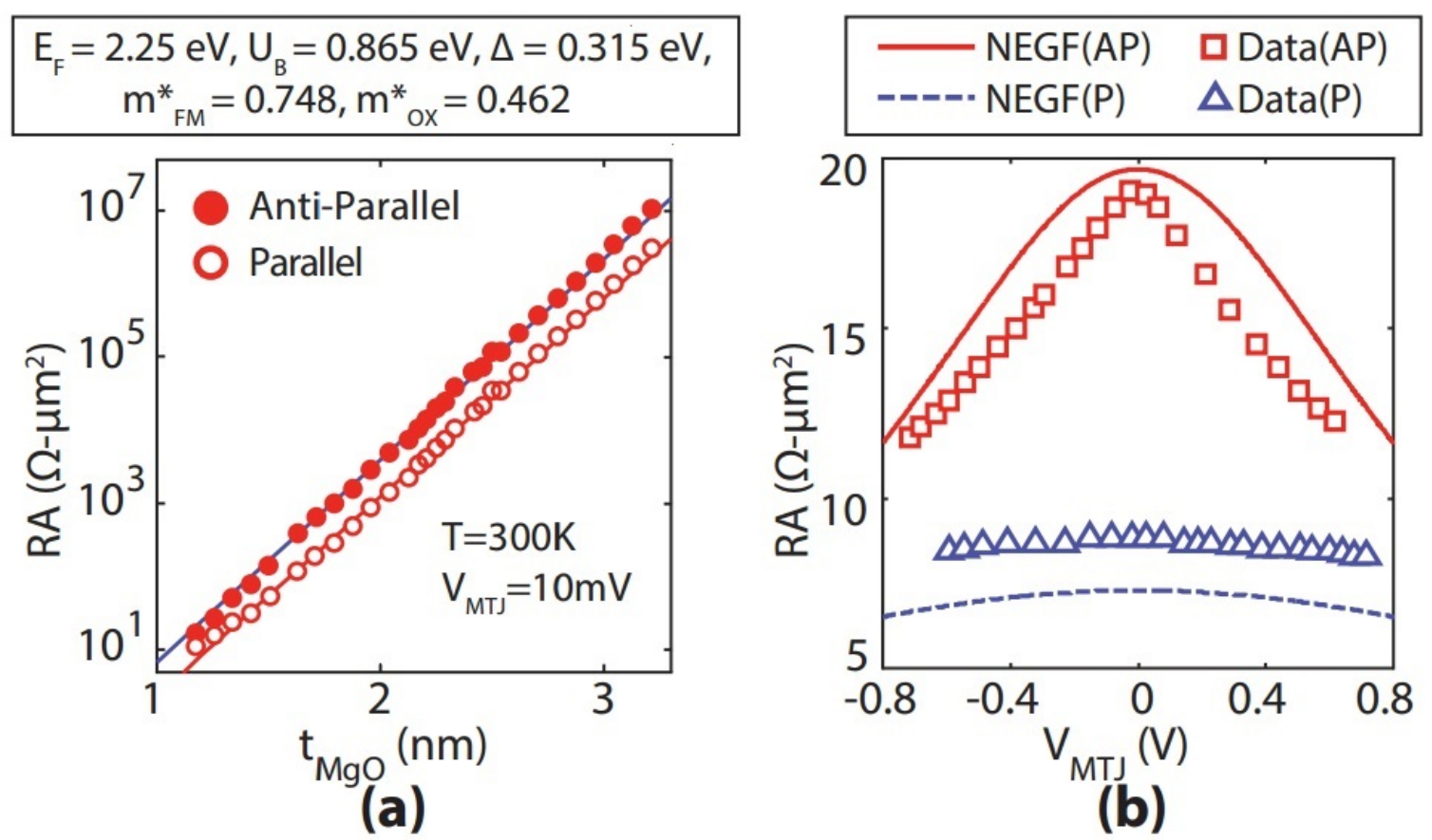}
\caption{The NEGF based transport simulation framework was calibrated to experimental results illustrated in \cite{yuasa2004giant,lin200945nm}. MTJ resistance varies with (a) oxide thickness and (b) applied voltage.}
\label{fig:calib_mtj}
\end{figure}

\begin{figure}
\centering
\includegraphics[width = 3.2in ]{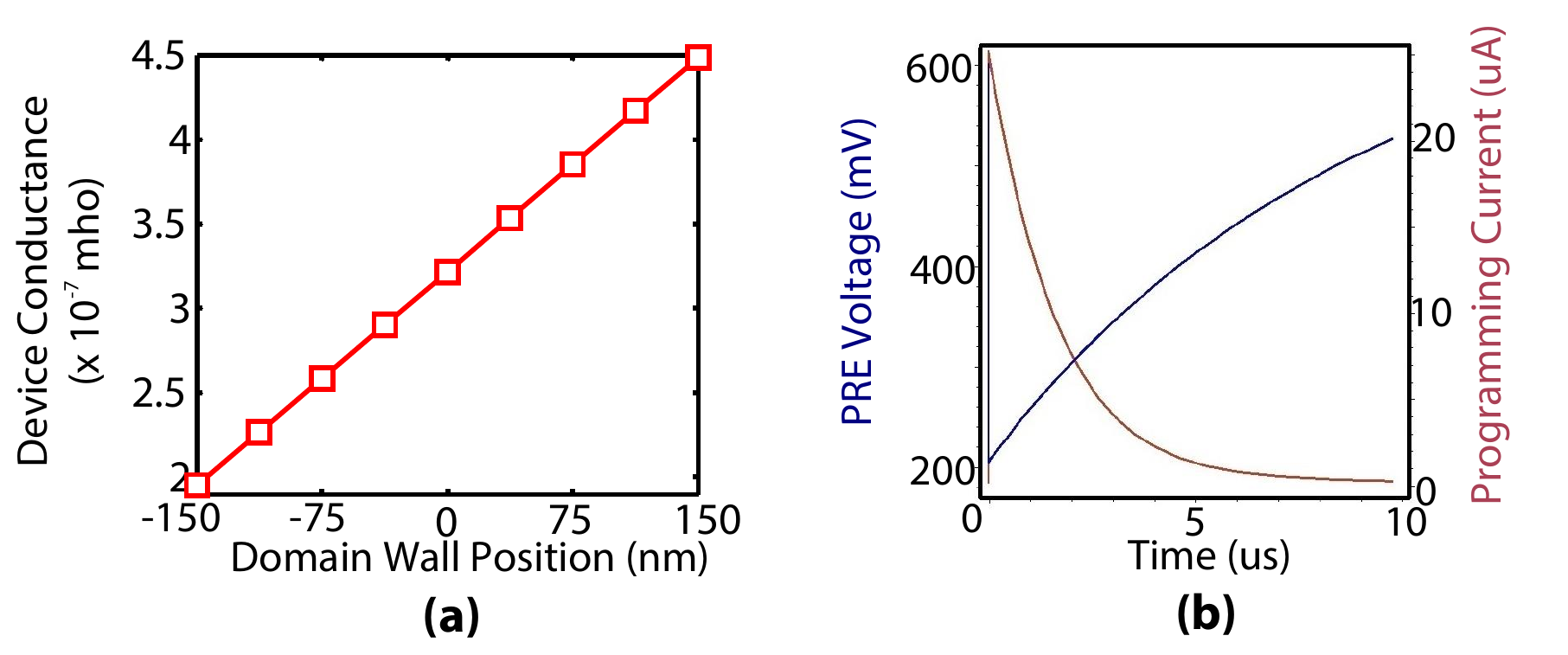}
\caption{(a) Linear variation of device conductance with domain wall position. (b) Programming circuit simulation to generate the STDP characteristics in the proposed spintronic synapse.}
\label{fig:negf_stdp}
\end{figure}

Non-Equilibrium Green's Function (NEGF) based transport simulation framework \cite{fong2011knack} was used to model the variation of the MTJ resistance with oxide thickness (Fig. \ref{fig:calib_mtj}(a)) and applied voltage (Fig. \ref{fig:calib_mtj}(b)) respectively. In order to determine the MTJ resistance for a FM with a domain wall separating two oppositely polarized magnetized domains, the NEGF based simulator \cite{fong2011knack} was modified by considering the parallel connection of three MTJs. The magnetization direction of the FL of the three MTJs were considered parallel, anti-parallel and perpendicular (domain wall) to the pinned layer magnetization. The length of the first two MTJs was varied according to the position of the domain wall while the width of the third MTJ was taken to be equal to the domain wall width. Fig. \ref{fig:negf_stdp}(a) depicts the variation of the device conductance with domain wall position (origin at the middle of the FM). In order to ensure proper synaptic functionality, it is also essential that the device resistance (for a particular position of the domain wall) does not vary with the voltage drop across the device. This is ensured by appropriately interfacing the device with the DPI circuit discussed earlier to generate the synaptic dynamics. The range of synapse resistances are in the $M\Omega$ range while the current flowing through the MTJ is in the range of a few $nA$s. Hence the voltage drop across the MTJ should be $\sim$ a few $mV$ ($<100mV$). It is apparent from Fig. \ref{fig:calib_mtj}(b) that the operating range of $V_{MTJ}$ is low enough to ensure negligible variation of the device conductance with device voltage drop for a particular domain wall position. As explained in the earlier section, such a linear variation of the device conductance with domain wall position results in the programming current being directly proportional to the relative conductance (weight) change involved. Hence the temporal profile of the necessary programming current also follows the STDP characteristics.

\subsection{Circuit Level Simulations}\label{subsec:circuit_simulation}

The programming and neuron circuits were simulated using a standard cell library in 45nm commercial CMOS technology. Although biological time scales are in the range of $\sim$ ms, it is not essential to limit the processing speed of the circuit to such slow time constants for implementing pattern recognition systems \cite{rajendran2013specifications}. The circuits were designed to operate at time constants in the range of $\sim \mu$s. 

Fig. \ref{fig:negf_stdp} (b) shows the response of the programming circuit for the case when the programming current path is active throughout the simulation time. The gate voltage of the transistor $M_{STDP}$ increases linearly and is reset at each input pulse leading to exponential sub-threshold current dynamics. The average power consumption of the circuit is 0.46$\mu W$ for the entire positive time window. The duration of the time window can be varied by changing the capacitance value. Further, this programming circuit can be shared by synapses in a particular column. It is worth noting here, that this power consumption does not include the power consumed in the $M_{STDP}$ transistor as current will flow through it only when the programming current path is activated for 1$ns$. The supply voltage for $M_{STDP}$ transistor was maintained at $600mV$ and hence the maximum amount of energy consumption involved in synapse programming is $\sim 48fJ (600mV \times 80\mu A \times 1 ns)$ per synaptic event.

\begin{figure*}
\centering
\includegraphics[width = 4.4in ]{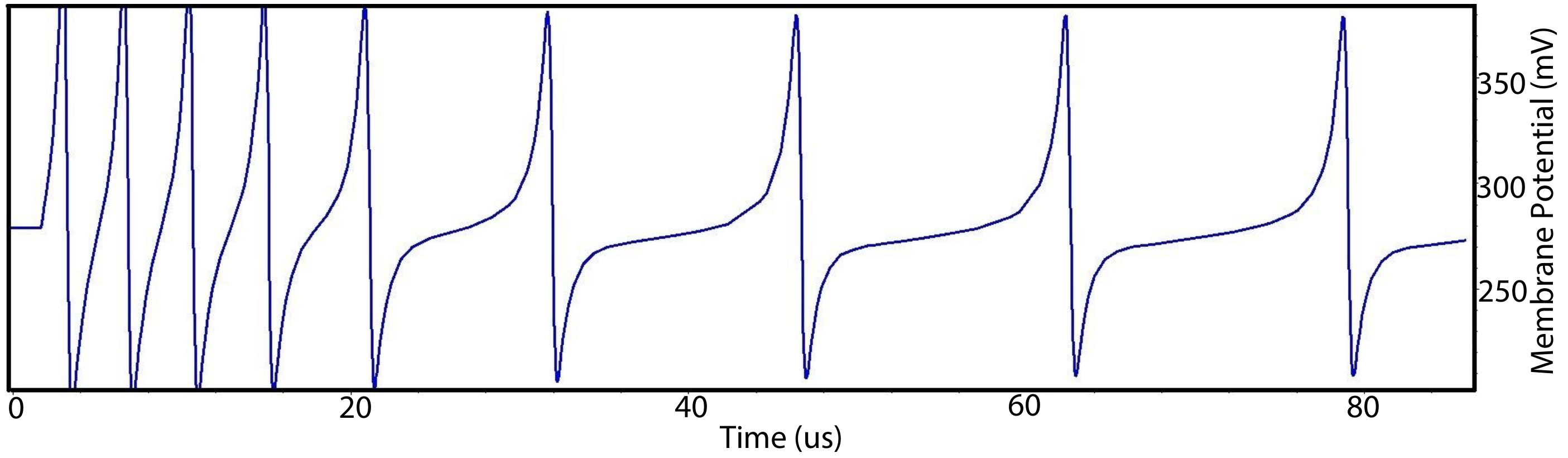}
\caption{CMOS neuron response to a constant input current with positive feedback, spike frequency adaptation and refractory period implementations.}
\label{fig:neuron_sim}
\end{figure*}

Fig. \ref{fig:neuron_sim} depicts the response of the CMOS neuron to a constant input current. As explained earlier, spike frequency adaptation scheme reduces the spike frequency to a steady state value. For a membrane capacitance of 50$fF$, the average power consumption of the circuit was $\sim$5.7$pJ$/ spike.

\subsection{System Level Simulations}\label{subsec:system_simulation}

\begin{figure*}
\centering
\includegraphics[width = 7in ]{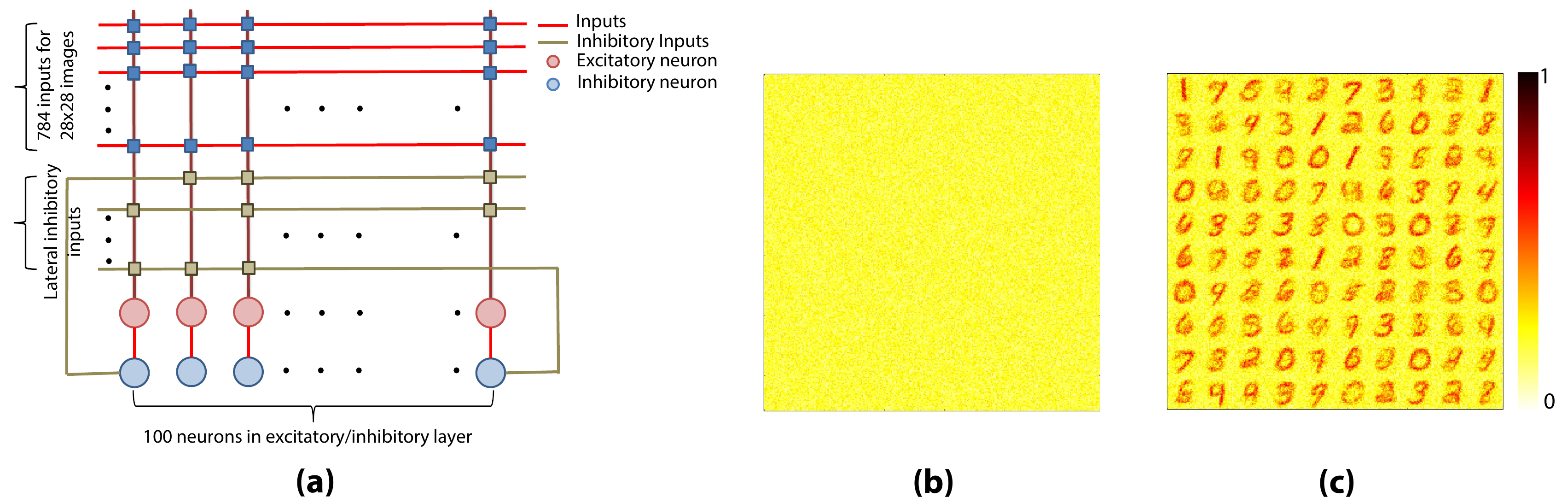}
\caption{(a) SNN topology used for digit recognition arranged in a crossbar array fashion. (b) Initial random synapse weights plotted in a $28 \times 28$ array for 100 neurons in the excitatory layer. (c) Representative digit patterns start getting stored in the synapse weights for each neuron after 1000 learning epochs.}
\label{fig:training}
\end{figure*}

The device and circuit behavioral models were used to simulate an SNN for digit recognition problems. The input images ($28 \times 28$ pixels) used for training was taken from the MNIST dataset \cite{lecun1998gradient}. The images were rate encoded and an array of 100 excitatory neurons was used to simulate the self-learning functionality of synapses in SNNs. Fig. \ref{fig:training} (a) demonstrates the SNN topology used for the recognition problem arranged in a crossbar array fashion. Synapses present at the crosspoints joining the inputs to the excitatory neurons can be programmed depending on the temporal spiking patterns of the pre- and post-neuron. Note that a synapse is absent at the crosspoint joining the excitatory to the inhibitory neuron. Inhibitory neurons are exactly similar to the excitatory neurons except that the output voltage spikes are negative.

Fig. \ref{fig:training} (b)-(c) depicts synapse weights plotted in $28 \times 28$ array (same as input images) for each of the 100 neurons used for the recognition purpose. Initially all the weights are random. However, as learning progresses the synapses of each neuron start learning generic representations of the various digits. Thus a particular neuron becomes more sensitive to the digit whose generic representation is being stored in its synapse weights since it will fire more if input spike trains are received at the pixel locations corresponding to high synaptic weights. The various system level simulation parameters have been outlined in Table \ref{table_2}. The parameters were tuned to achieve learning ability in the synapses. The units of the time constants are with respect to the duration of each timestep in the simulation. For this work, the circuits were designed to operate in $\sim \mu s$ time scale as mentioned before. It is worth noting here that the manner in which the time constants and other parameters can be tuned in the circuit level simulations have been discussed in the previous section. The numbers in braces represent the value corresponding to the inhibitory neuron. 
\begin{table}[h]
\renewcommand{\arraystretch}{1.3}
\caption{System Simulation Parameters}
\label{table_2}
\centering
\begin{tabular}{c c}
\hline 
\bfseries { Parameters} & \bfseries { Value}\\
\hline
{No. of excitatory/inhibitory neurons} & { $100$} \\
{Probability of input spike per timestep} & { $ 0-0.06375$} \\
%{Membrane potential range } & { $ 23 (5) mV$} \\
{Number of timesteps per image} & { $350$} \\
{STDP time constants} & { $100 (1)$ }\\
{Neuron time constants} & {$10 (10)$} \\
{ Post-synaptic current time constants} & { 1 (2)} \\
\hline
\end{tabular}
\end{table}

Additionally, we would like to mention here, that such neuromorphic systems are significantly robust to imprecision due to device mismatch, variability and noise effects due to the adaptive nature of such computations involving plasticity, homeostasis and feedback mechanisms \cite{chicca2014neuromorphic}. Further, authors in Ref. \cite{querlioz2013immunity} demonstrate the immunity of such single layer SNNs based on crossbar arrays of resistive synapses with lateral inhibition and homeostasis effects to variations and non-idealities in typical resistive synaptic devices and CMOS neuron circuits. In particular, we performed an analysis of the impact of variations in the oxide thickness/MTJ synaptic conductances on the classification accuracy of the system. Almost no degradation in classification accuracy was observed for the 100-neuron network even with $25\%$ variation in the resistances of the spintronic synapses.
\begin{table*}[t]
\scriptsize
\renewcommand{\arraystretch}{1.3}
\caption{Comparison with other proposed synapses}
\label{table_3}
\centering
\begin{tabular}{ p{3cm} p{2.8cm} p{3.2cm} p{2.2cm} p{1.5cm} p{3.6cm}}
\hline 
\hline
\bfseries {Device} & \bfseries {Dimensions} & \bfseries {Programming Energy/ Operating Voltage} & \bfseries {Programming Time} & \bfseries {Terminals} & \bfseries {Programming Mechanism}\\
\hline
{GeSbTe memristor \cite{jackson2013nanoscale}} & {$40nm$ mushroom and $10nm$ pore} & {Average $2.74pJ$/ event} & {$60ns$} & {$2$} & {Programmed by Joule heating (Phase change)}\\
{GeSbTe memristor \cite{kuzum2011nanoelectronic}} & {$75nm$ electrode diameter} & {$50pJ$ (reset) \& $0.675pJ$ (set)} & {$10ns$} & {$2$} & {Programmed by Joule heating (Phase change)}\\
{Ag/AgInSbTe/Ag chalcogenide memristor \cite{li2014activity}} & {$100\mu m$ x $100\mu m$} & {Threshold voltage - $0.3V$} & {$5\mu s$} & {$2$} & {Programmed by Joule heating (Phase change)}\\
{Ag-Si memristor \cite{jo2010nanoscale}} & {$100nm$ x $100nm$} & {Threshold voltage - $2.2V$} & {$300\mu s$} & {$2$} & {Movement of Ag ions}\\
{FeFET \cite{nishitani2013dynamic}} & {Channel length - $3\mu m$} & {Maximum gate voltage - $4V$} & {$ 10 \mu s$} & {$3$} & {Gate voltage modulation of ferroelectric polarization}\\
{Floating gate transistor \cite{ramakrishnan2011floating}} & {$1.8\mu m/0.6\mu m (0.35\mu m$ CMOS technology)} & {$V_{dd} - 4.2V$ \& Tunneling Voltage -$15V$} & {$100\mu s$ (injection) \& $2ms$ (tunneling)} & {$3$} & {Injection and tunneling currents}\\
{SRAM synapse \cite{rajendran2013specifications}} & {$0.3\mu m^2$ ($10nm$ CMOS technology)} & {Average $328fJ$ (4-bit synapse)} & {-} & {-} & {Digital counter based circuits}\\
{Spintronic synapse} & {Ferromagnet dimensions - $320nm$ x $20nm$} & {Maximum $48fJ$/ event} & {$1ns$} & {$3$} & {Spin-orbit torque}\\
\hline
\hline
\end{tabular}
\end{table*}

\section{Conclusions}\label{sec:discussions}

While prior proposals have investigated mono-domain spintronic devices for implementing spiking neurons \cite{sengupta2015magnetic} and short-term plasticity effects \cite{sengupta2016short}, to the best of our knowledge this is the first work to propose a hybrid spintronic-CMOS SNN design with self-learning (from the device to the system level) based on a three-terminal multi-domain spintronic synapse device structure consisting of decoupled spike transmission and programming current paths. This is advantageous for implementation of neuromorphic systems capable of on-chip learning since the programming current path is independent of the read current path. Interface CMOS circuit design for self-learning is highly simplified since the resistance in the programming current path is constant and determined mainly by the HM resistance and independent of the synapse conductance. 

Table \ref{table_3} provides a comparative analysis of our spintronic synapse (calibrated to experiments performed in Ref. \cite{emori2014spin}) with other proposed synaptic devices. Synaptic device structures based on emerging post-CMOS technologies \cite{jackson2013nanoscale,jo2010nanoscale,kuzum2011nanoelectronic} are usually two-terminal devices and do not offer de-coupled programming and read current paths. Additionally they are usually characterized by relatively high programming energies. In contrast, our proposed synapse offers low programming energy and requires very small programming time. A maximum programming energy of $\sim 48fJ$ is consumed per synaptic event due to the highly energy-efficient spin-orbit torque induced synaptic plasticity. Three terminal synaptic devices based on FeFET \cite{nishitani2013dynamic} and floating gate transistors \cite{ramakrishnan2011floating} have been also proposed. However, the programming in such devices is usually accomplished through the gate terminal and a high gate voltage is usually applied across a very thin oxide \cite{ramakrishnan2011floating,nishitani2013dynamic} leading to reliability issues, in addition to associated high power consumption. Programming is also relatively slow in such three terminal synaptic devices \cite{ramakrishnan2011floating,nishitani2013dynamic}. It is worth noting here, that the current flowing through the oxide in the MTJ structure for our proposed synapse is the read current which is $\sim nA$ and drives sub-threshold CMOS circuits. SRAM based synapses have been also proposed for digital CMOS based SNN design \cite{rajendran2013specifications}. However, for implementing 1 bit of the synapse, an 8-T SRAM cell has to be used, thereby leading to significant area overhead for implementation of a single synapse \cite{rajendran2013specifications}. In addition, learning circuits will involve multiple digital counters and will be more area/power consuming than our proposed design.

Interested readers are referred to Ref. \cite{noguchi20157} for a discussion on the practical implementation of arrays of such spintronic devices interfaced with CMOS transistors. The size limitation of crossbar arrays of such spintronic devices is determined by the driving capabilities of rows of the array by input voltages in the presence of parasitics. In addition, sneak paths also become a potential issue for large crossbar arrays in order to implement on-chip learning. These are concerns that are equally valid for spin-devices and other memristive technologies, in general. However, it is worth noting here that computation occurring in a large crossbar can be distributed easily among smaller crossbar arrays by simply replacing the large unit by an equivalent number of smaller crossbar units using peripheral control circuitry. 

In conclusion, we formulated a device, circuit and algorithm co-simulation framework calibrated to experimental results to validate the functionalities and performance of the proposed hybrid spintronic-CMOS based SNN design with on-chip learning. We proposed circuit primitives for generating STDP in the proposed synapse and demonstrated how such synaptic devices could be arranged in a crossbar fashion leading to an area and power efficient SNN implementation that is capable of recognizing patterns in input data. Simulation studies indicate the efficiency of the proposed hybrid spintronic-CMOS based SNN design as an ultra-low power neuromorphic computing platform capable of online learning.

\section*{Acknowledgment}

The work was supported in part by, Center for Spintronic Materials, Interfaces, and Novel Architectures (C-SPIN), a MARCO and DARPA sponsored StarNet center, by the Semiconductor
Research Corporation, the National Science Foundation, Intel Corporation and by the National Security Science and Engineering Faculty Fellowship.


\begin{thebibliography}{10}
\providecommand{\url}[1]{#1}
\csname url@samestyle\endcsname
\providecommand{\newblock}{\relax}
\providecommand{\bibinfo}[2]{#2}
\providecommand{\BIBentrySTDinterwordspacing}{\spaceskip=0pt\relax}
\providecommand{\BIBentryALTinterwordstretchfactor}{4}
\providecommand{\BIBentryALTinterwordspacing}{\spaceskip=\fontdimen2\font plus
\BIBentryALTinterwordstretchfactor\fontdimen3\font minus
\fontdimen4\font\relax}
\providecommand{\BIBforeignlanguage}[2]{{%
\expandafter\ifx\csname l@#1\endcsname\relax
\typeout{** WARNING: IEEEtran.bst: No hyphenation pattern has been}%
\typeout{** loaded for the language `#1'. Using the pattern for}%
\typeout{** the default language instead.}%
\else
\language=\csname l@#1\endcsname
\fi
#2}}
\providecommand{\BIBdecl}{\relax}
\BIBdecl


\bibitem{ghosh2009spiking}
S.~Ghosh-Dastidar and H.~Adeli, ``Spiking neural networks,''
  \emph{International journal of neural systems}, vol.~19, no.~04, pp.
  295--308, 2009.

\bibitem{markram2006blue}
H.~Markram, ``The blue brain project,'' \emph{Nature Reviews Neuroscience},
  vol.~7, no.~2, pp. 153--160, 2006.

\bibitem{schemmel2008wafer}
J.~Schemmel, J.~Fieres, and K.~Meier, ``Wafer-scale integration of analog
  neural networks,'' in \emph{Neural Networks, 2008. IJCNN 2008.(IEEE World
  Congress on Computational Intelligence). IEEE International Joint Conference
  on}.\hskip 1em plus 0.5em minus 0.4em\relax IEEE, 2008, pp. 431--438.

\bibitem{jin2010modeling}
X.~Jin, M.~Lujan, L.~A. Plana, S.~Davies, S.~Temple, and S.~Furber, ``Modeling
  spiking neural networks on {S}pi{N}{N}aker,'' \emph{Computing in Science \&
  Engineering}, vol.~12, no.~5, pp. 91--97, 2010.

\bibitem{merolla2011digital}
P.~A. Merolla, J.~V. Arthur, R.~Alvarez-Icaza, A.~S. Cassidy, J.~Sawada,
  F.~Akopyan, B.~L. Jackson, N.~Imam, C.~Guo, Y.~Nakamura \emph{et~al.}, ``A
  million spiking-neuron integrated circuit with a scalable communication
  network and interface,'' \emph{Science}, vol. 345, no. 6197, pp. 668--673,
  2014.

\bibitem{rajendran2013specifications}
B.~Rajendran, Y.~Liu, J.-s. Seo, K.~Gopalakrishnan, L.~Chang, D.~J. Friedman,
  and M.~B. Ritter, ``Specifications of nanoscale devices and circuits for
  neuromorphic computational systems,'' \emph{Electron Devices, IEEE
  Transactions on}, vol.~60, no.~1, pp. 246--253, 2013.

\bibitem{jackson2013nanoscale}
B.~L. Jackson, B.~Rajendran, G.~S. Corrado, M.~Breitwisch, G.~W. Burr,
  R.~Cheek, K.~Gopalakrishnan, S.~Raoux, C.~T. Rettner, A.~Padilla
  \emph{et~al.}, ``Nanoscale electronic synapses using phase change devices,''
  \emph{ACM Journal on Emerging Technologies in Computing Systems (JETC)},
  vol.~9, no.~2, p.~12, 2013.

\bibitem{jo2010nanoscale}
S.~H. Jo, T.~Chang, I.~Ebong, B.~B. Bhadviya, P.~Mazumder, and W.~Lu,
  ``Nanoscale memristor device as synapse in neuromorphic systems,'' \emph{Nano
  letters}, vol.~10, no.~4, pp. 1297--1301, 2010.

\bibitem{ramakrishnan2011floating}
S.~Ramakrishnan, P.~E. Hasler, and C.~Gordon, ``Floating gate synapses with
  spike-time-dependent plasticity,'' \emph{Biomedical Circuits and Systems,
  IEEE Transactions on}, vol.~5, no.~3, pp. 244--252, 2011.

\bibitem{nishitani2013dynamic}
Y.~Nishitani, Y.~Kaneko, M.~Ueda, E.~Fujii, and A.~Tsujimura, ``Dynamic
  observation of brain-like learning in a ferroelectric synapse device,''
  \emph{Japanese Journal of Applied Physics}, vol.~52, no.~4S, p. 04CE06, 2013.

\bibitem{kuzum2011nanoelectronic}
D.~Kuzum, R.~G. Jeyasingh, B.~Lee, and H.-S.~P. Wong, ``Nanoelectronic
  programmable synapses based on phase change materials for brain-inspired
  computing,'' \emph{Nano letters}, vol.~12, no.~5, pp. 2179--2186, 2011.

\bibitem{sharad2012spin}
M.~Sharad, C.~Augustine, G.~Panagopoulos, and K.~Roy, ``Spin-based neuron model
  with domain-wall magnets as synapse,'' \emph{Nanotechnology, IEEE
  Transactions on}, vol.~11, no.~4, pp. 843--853, 2012.

\bibitem{ramasubramanian2014spindle}
S.~G. Ramasubramanian, R.~Venkatesan, M.~Sharad, K.~Roy, and A.~Raghunathan,
  ``{S}{P}{I}{N}{D}{L}{E}: {S}{P}{I}{N}tronic deep learning engine for
  large-scale neuromorphic computing,'' in \emph{Proceedings of the 2014
  international symposium on Low power electronics and design}.\hskip 1em plus
  0.5em minus 0.4em\relax ACM, 2014, pp. 15--20.

\bibitem{:/content/aip/journal/apl/106/14/10.1063/1.4917011}
\BIBentryALTinterwordspacing
A.~Sengupta, S.~H. Choday, Y.~Kim, and K.~Roy, ``Spin orbit torque based
  electronic neuron,'' \emph{Applied Physics Letters}, vol. 106, no.~14,
  pp.~--, 2015. [Online]. Available:
  \url{http://scitation.aip.org/content/aip/journal/apl/106/14/10.1063/1.4917011}
\BIBentrySTDinterwordspacing

\bibitem{morris1999hebb}
R.~Morris, ``The organization of behavior, {W}iley: New york; 1949,''
  \emph{Brain research bulletin}, vol.~50, no.~5, p. 437, 1999.

\bibitem{bi2001synaptic}
G.-q. Bi and M.-m. Poo, ``Synaptic modification by correlated activity:
  {H}ebb's postulate revisited,'' \emph{Annual review of neuroscience},
  vol.~24, no.~1, pp. 139--166, 2001.

\bibitem{peter2015unsup}
P.~U. Diehl and M.~Cook, ``Unsupervised learning of digit recognition using
  spike-timing-dependent plasticity,'' \emph{Frontiers in Computational
  Neuroscience}, vol.~9, p.~99, 2015.

\bibitem{knagsparse}
P.~Knag, J.~K. Kim, T.~Chen, and Z.~Zhang, ``A sparse coding neural network
  asic with on-chip learning for feature extraction and encoding,''
  \emph{Solid-State Circuits, IEEE Journal of}, vol.~50, no.~4, pp. 1070--1079,
  2015.

\bibitem{emori2013current}
S.~Emori, U.~Bauer, S.-M. Ahn, E.~Martinez, and G.~S. Beach, ``Current-driven
  dynamics of chiral ferromagnetic domain walls,'' \emph{Nature materials},
  vol.~12, no.~7, pp. 611--616, 2013.

\bibitem{chen2013novel}
G.~Chen, J.~Zhu, A.~Quesada, J.~Li, A.~N’Diaye, Y.~Huo, T.~Ma, Y.~Chen,
  H.~Kwon, C.~Won \emph{et~al.}, ``Novel chiral magnetic domain wall structure
  in {Fe}/{Ni}/{Cu} (001) films,'' \emph{Physical review letters}, vol. 110,
  no.~17, p. 177204, 2013.

\bibitem{martinez2014current}
E.~Martinez, S.~Emori, N.~Perez, L.~Torres, and G.~S. Beach, ``Current-driven
  dynamics of {Dzyaloshinskii} domain walls in the presence of in-plane fields:
  {Full} micromagnetic and one-dimensional analysis,'' \emph{Journal of Applied
  Physics}, vol. 115, no.~21, p. 213909, 2014.

\bibitem{emori2014spin}
S.~Emori, E.~Martinez, K.-J. Lee, H.-W. Lee, U.~Bauer, S.-M. Ahn, P.~Agrawal,
  D.~C. Bono, and G.~S. Beach, ``Spin {Hall} torque magnetometry of
  {Dzyaloshinskii} domain walls,'' \emph{Physical Review B}, vol.~90, no.~18,
  p. 184427, 2014.

\bibitem{perez2014micromagnetic}
N.~Perez, L.~Torres, and E.~Martinez-Vecino, ``Micromagnetic {Modeling} of
  {Dzyaloshinskii}--{Moriya} {Interaction} in {Spin} {Hall} {Effect}
  {Switching},'' \emph{Magnetics, IEEE Transactions on}, vol.~50, no.~11, pp.
  1--4, 2014.

\bibitem{hirsch1999spin}
J.~Hirsch, ``Spin hall effect,'' \emph{Physical Review Letters}, vol.~83,
  no.~9, p. 1834, 1999.

\bibitem{sengupta2016spin}
A.~Sengupta, Z.~Al~Azim, X.~Fong, and K.~Roy, ``Spin-orbit torque induced
  spike-timing dependent plasticity,'' \emph{Applied Physics Letters}, vol.
  106, no.~9, p. 093704, 2015.

\bibitem{lazzaro1994low}
J.~Lazzaro and J.~Wawrzynek, \emph{Low-power silicon neurons, axons and
  synapses}.\hskip 1em plus 0.5em minus 0.4em\relax Springer, 1994.

\bibitem{bartolozzi2007synaptic}
C.~Bartolozzi and G.~Indiveri, ``Synaptic dynamics in analog {V}{L}{S}{I},''
  \emph{Neural computation}, vol.~19, no.~10, pp. 2581--2603, 2007.

\bibitem{chicca2014neuromorphic}
E.~Chicca, F.~Stefanini, C.~Bartolozzi, and G.~Indiveri, ``Neuromorphic
  electronic circuits for building autonomous cognitive systems,''
  \emph{Proceedings of the IEEE}, vol. 102, no.~9, pp. 1367--1388, 2014.

\bibitem{indiveri2013integration}
G.~Indiveri, R.~Legenstein, G.~Deligeorgis, T.~Prodromakis \emph{et~al.},
  ``Integration of nanoscale memristor synapses in neuromorphic computing
  architectures,'' \emph{Nanotechnology}, vol.~24, no.~38, p. 384010, 2013.

\bibitem{indiveri2003low}
G.~Indiveri, ``A low-power adaptive integrate-and-fire neuron circuit,'' in
  \emph{ISCAS (4)}, 2003, pp. 820--823.

\bibitem{livi2009current}
P.~Livi and G.~Indiveri, ``A current-mode conductance-based silicon neuron for
  address-event neuromorphic systems,'' in \emph{Circuits and systems, 2009.
  ISCAS 2009. IEEE international symposium on}.\hskip 1em plus 0.5em minus
  0.4em\relax IEEE, 2009, pp. 2898--2901.

\bibitem{vansteenkiste2011mumax}
A.~Vansteenkiste, J.~Leliaert, M.~Dvornik, M.~Helsen, F.~Garcia-Sanchez, and
  B.~Van~Waeyenberge, ``The design and verification of mumax3,'' \emph{AIP
  Advances}, vol.~4, no.~10, p. 107133, 2014.

\bibitem{goodman2009brian}
D.~F. Goodman and R.~Brette, ``The brian simulator,'' \emph{Frontiers in
  neuroscience}, vol.~3, no.~2, p. 192, 2009.

\bibitem{slonczewski1989conductance}
J.~C. Slonczewski, ``Conductance and exchange coupling of two ferromagnets
  separated by a tunneling barrier,'' \emph{Physical Review B}, vol.~39,
  no.~10, p. 6995, 1989.

\bibitem{yuasa2004giant}
S.~Yuasa, T.~Nagahama, A.~Fukushima, Y.~Suzuki, and K.~Ando, ``Giant
  room-temperature magnetoresistance in single-crystal {Fe}/{MgO}/{Fe} magnetic
  tunnel junctions,'' \emph{Nature materials}, vol.~3, no.~12, pp. 868--871,
  2004.

\bibitem{lin200945nm}
C.~Lin, S.~Kang, Y.~Wang, K.~Lee, X.~Zhu, W.~Chen, X.~Li, W.~Hsu, Y.~Kao,
  M.~Liu \emph{et~al.}, ``45nm low power {C}{M}{O}{S} logic compatible embedded
  {S}{T}{T} {M}{R}{A}{M} utilizing a reverse-connection 1{T}/1{M}{T}{J} cell,''
  in \emph{Electron Devices Meeting (IEDM), 2009 IEEE International}.\hskip 1em
  plus 0.5em minus 0.4em\relax IEEE, 2009, pp. 1--4.

\bibitem{fong2011knack}
X.~Fong, S.~K. Gupta, N.~N. Mojumder, S.~H. Choday, C.~Augustine, and K.~Roy,
  ``K{N}{A}{C}{K}: {A} hybrid spin-charge mixed-mode simulator for evaluating
  different genres of spin-transfer torque {M}{R}{A}{M} bit-cells,'' in
  \emph{Simulation of Semiconductor Processes and Devices (SISPAD), 2011
  International Conference on}.\hskip 1em plus 0.5em minus 0.4em\relax IEEE,
  2011, pp. 51--54.

\bibitem{lecun1998gradient}
Y.~LeCun, L.~Bottou, Y.~Bengio, and P.~Haffner, ``Gradient-based learning
  applied to document recognition,'' \emph{Proceedings of the IEEE}, vol.~86,
  no.~11, pp. 2278--2324, 1998.

\bibitem{querlioz2013immunity}
D.~Querlioz, O.~Bichler, P.~Dollfus, and C.~Gamrat, ``Immunity to device
  variations in a spiking neural network with memristive nanodevices,''
  \emph{Nanotechnology, IEEE Transactions on}, vol.~12, no.~3, pp. 288--295,
  2013.

\bibitem{li2014activity}
Y.~Li, Y.~Zhong, J.~Zhang, L.~Xu, Q.~Wang, H.~Sun, H.~Tong, X.~Cheng, and
  X.~Miao, ``Activity-dependent synaptic plasticity of a chalcogenide
  electronic synapse for neuromorphic systems,'' \emph{Scientific reports},
  vol.~4, 2014.

\bibitem{sengupta2015magnetic}
A.~Sengupta, P.~Panda, P.~Wijesinghe, Y.~Kim, and K.~Roy, ``Magnetic tunnel
  junction mimics stochastic cortical spiking neurons,'' \emph{Scientific
  reports}, vol.~6, p. 30039, 2016.

\bibitem{sengupta2016short}
A.~Sengupta and K.~Roy, ``Short-term plasticity and long-term potentiation in
  magnetic tunnel junctions: Towards volatile synapses,'' \emph{Physical Review
  Applied}, vol.~5, no.~2, p. 024012, 2016.

\bibitem{noguchi20157}
H.~Noguchi, K.~Ikegami, K.~Kushida, K.~Abe, S.~Itai, S.~Takaya, N.~Shimomura,
  J.~Ito, A.~Kawasumi, H.~Hara \emph{et~al.}, ``7.5 a 3.3 ns-access-time
  71.2$\mu$w/mhz 1mb embedded stt-mram using physically eliminated read-disturb
  scheme and normally-off memory architecture,'' in \emph{2015 IEEE
  International Solid-State Circuits Conference-(ISSCC) Digest of Technical
  Papers}.\hskip 1em plus 0.5em minus 0.4em\relax IEEE, 2015, pp. 1--3.

\end{thebibliography}
\end{document}